\documentclass[graybox]{svmult}

\usepackage{helvet}         
\usepackage{courier}        
\usepackage{type1cm}        
\usepackage{makeidx}         
\usepackage{graphicx}        
\usepackage{multicol}        
\usepackage[bottom]{footmisc}

\makeindex             

\begin{document}

\title*{Modeling pedestrian evacuation movement in a swaying ship}
\author{Juan Chen, Jian Ma and SM Lo}
\institute{Juan Chen and SM Lo \at Department of Architecture and Civil Engineering, City University of Hong Kong, Hong Kong SAR, China\\ \email{juanchen6-c@my.cityu.edu.hk; bcsmli@cityu.edu.hk}
\and Jian Ma (Corresponding author)\at School of Transportation and Logistics, Southwest Jiaotong University, Chengdu, 610031, China; \\National United Engineering Laboratory of Integrated and Intelligent Transportation, Southwest Jiaotong University, Chengdu, 610031, China\\
 \email{majian@mail.ustc.edu.cn}
 }
%
%
\maketitle

\abstract{With the advance in living standard, cruise travel has been rapidly expanding around the world in recent years. The transportation of passengers in water has also made a rapid development. It is expected that ships will be more and more widely used. Unfortunately, ship disasters occurred in these years caused serious losses. It raised the concern on effectiveness of passenger evacuation on ships. The present study thus focuses on pedestrian evacuation features on ships. On ships, passenger movements are affected by the periodical water motion and thus are quite different from the characteristic when walking on static horizontal floor. Taking into consideration of this special feature, an agent-based pedestrian model is formulized and the effect of ship swaying on pedestrian evacuation efficiency is investigated. Results indicated that the proposed model can be used to quantify the special evacuation process on ships.}

\section{Introduction}
\label{sec:1}
In the past decades, cruise travel has been a rapid expanding field around the world. A worldwide annual growth rate of 6.55\% for passage by sea from 1990 to 2019 has been recorded which is expected to grow continuously. The transportation of passengers in water has also made a rapid development. It is expected that ships will be more and more widely used in transporting passengers. Unfortunately, several ship disasters occurred in these years which caused serious losses to people's lives and properties. It raised the concern on the effectiveness of large crowd passenger evacuation on ships.

To mitigate the ill effect of an emergency, evacuating passengers from the ships will be necessary. The International Maritime Organization (IMO) has already developed guidelines for passenger ship evacuation. However, these guidelines and regulations, like most building codes, only enforces capacity limit of individual components such as exits, passenger way width, etc., thus can barely provide efficient setting and management strategies. Computer simulation, on the contrary, can be helpful by performing simulations even at the ship design stage. The International Conventions for the Safety of Life at Sea (SOLAS) now requires evacuation analysis at the early stage of ship design (IMO MSC.1/Circ.1238, 2007). Thus some simulation models have been built to perform computer simulation based evacuation analysis.

When compared with building evacuation, passenger ship evacuation is still a new research topic and only limited publications can be found. Existing works were more or less inspired by the IMO guidelines. The two full scale drill exercise projects, i.e., "FIRE-EXIT" and "SAFEGUARD" have provided valuable information for model validation, especially for "maritime-Exodus" \cite{Galea2013}. Some other models have also been established including, AENEAS \cite{Valanto2006}, EVI \cite{Azzi} and VELOS \cite{Ginnis2010}. The former two models are grid-based models, and thus have great advantage in computing efficiency, yet the discretization may affect their ability to model environmental space and detailed pedestrian motion when compared with the last two continuous space models. The influences of ship motion on pedestrian movement in these models are almost mimicked by speed reduction considering different inclination angle rather than considering the coupled-forced pedestrian movement features.

Thus, in the present paper we establish an evacuation model taking into account forced pedestrian movement pattern. The rest of the present paper is organized as follows. In section 2, we introduce the model which takes into account ship swaying effect. In section 3, we firstly compare several single pedestrian movement features under different conditions and then analyze a ship evacuation process. In the last section, conclusions are made.

\section{Ship evacuation model}
\label{sec:2}
Basically, the design layout of the cruise ships is same as that of hotels on the ground, which includes separated accommodation cabins, restaurants, sports facilities, etc., so that cruise ships can be regarded as mobile hotels. However, one of the important differences between them is that the deck of cruise ship has a complex motion pattern as a result of periodical water movement. Thus pedestrian on board would be affected by the inertial force, which further makes pedestrian movement characteristic different from when walking on a static horizontal floor. Taking into consideration of this special feature, an agent-based pedestrian model is formulized, and the effect of ship swaying on pedestrian evacuation efficiency is investigated.

\subsection{Ship swaying features}
\label{subsec:1}
\begin{figure}[b]
\sidecaption
\includegraphics[bb=0 270 580 540, scale=0.3]{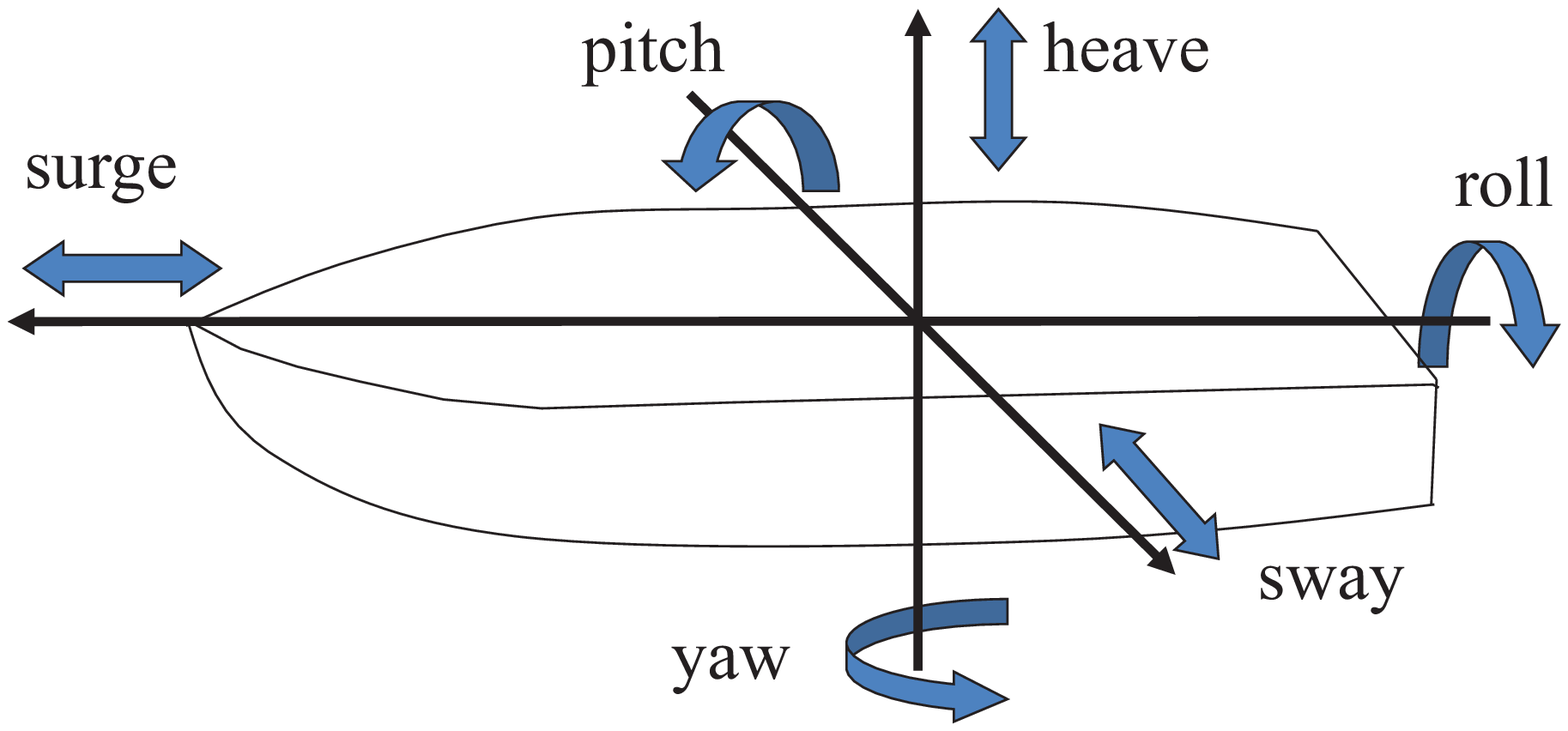}
\includegraphics[bb=0 290 280 540, scale=0.5]{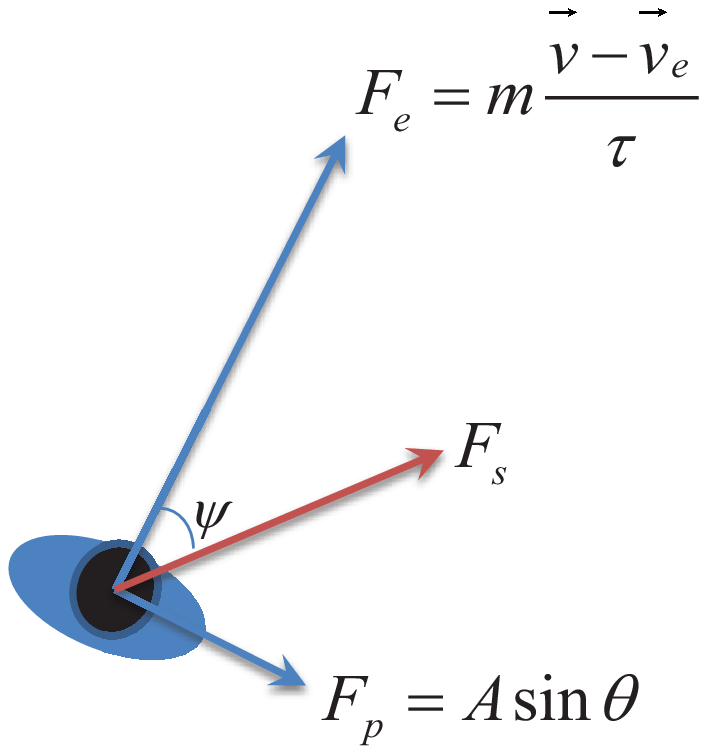}
%
%
\caption{(a) Scheme of ship motion types and (b) Force analysis for an on-board passenger.}
\label{fig:1}       
\end{figure}

As mentioned before, ship movement in water is very complex. It displays a six-degree-of-freedom motion feature because of the periodical water movement and wind influence, as shown in Fig.~\ref{fig:1}a. These motions include linear translation motions (surge, sway and heave) and nonlinear rotation motions (pitch, roll and yaw). Swaying refers to the linear side-to-side motion, while rolling and pitching represent the tilting rotation of a ship about its front-back axis and side-to-side axis, respectively. Due to combined motions of swaying, rolling and pitching, the deck of a cruise ship may get inclining and then recovering periodically as a result of the counter-rotating torque. Hereinafter, we do not distinguish these motion types and use swaying instead. This special combined movement feature can be quantified by two parameters, swaying amplitude $B(t)$ and phase $\varphi(t)$. It should be noticed that here $\varphi(t)$ is periodically changing with time $t$. The exact form of ship motions in seaway should base on water movement and wind influence features, however, for simplicity and without loss of generality, we assume,
\begin{eqnarray}
\varphi(t)= M \times sin(t),
\label{eq:01}
\end{eqnarray}
where $M$ means the maximal ship ground inclining degree.

\subsection{Pedestrian movement features}
\label{subsec:2}
When a pedestrian moves smoothly on a horizontal floor and there is no swaying, the pedestrian movement is driven by the internal self-driven force, $F_e$, as shown in Fig.~\ref{fig:1}b. This internal driven force makes the pedestrian move with a speed not far away from his expected speed $v_e$, thus the required evacuation time can be estimated based on walking speed, queuing length, and emergency exit capacity \cite{Daamen2012}. However, it should be noticed that the movement of a pedestrian is accomplished step by step using their two legs. As a consequence, there is a gait cycle, as reported in Ref\cite{Olivier2012,Ma2013}. Thus, when the ground is swaying, the pedestrian gait would be affected by the new force component, the inertial force $F_s$, as shown in Fig.~\ref{fig:1}b. This new force component can be projected onto two directions, i.e., the pedestrian movement direction and the direction perpendicular to his movement direction. The force component along pedestrian movement direction, $F_{e||}$, would affect the maximal speed that pedestrian can archive, while the force component along the lateral direction would affect the gait cycle, so we introduce a periodically changing force component along the lateral direction, i.e., $F_p$ to quantify this influence.

For the internal self-driven force, as in other force-based models \cite{Helbing2000,Chraibi2010}, we take the following form,
\begin{eqnarray}
F_e = m \frac{\vec{v} - \vec{v_e}}{\tau}
\label{eq:02}
\end{eqnarray}
where $\tau=0.5$s is selected according to Ref\cite{Ma2010}. The ship swaying induced parallel force component $F_{e||}$ can be denoted as,
\begin{eqnarray}
F_{e||} = F_s \times cos\psi
\label{eq:03}
\end{eqnarray}
Where $\psi$ means the angle between $F_s$ and $F_e$. For the step by step lateral movement of the pedestrian, we assume that,
\begin{eqnarray}
F_p = A \times sin\theta
\label{eq:04}
\end{eqnarray}
It should be noticed that a pedestrian can sense the ship swaying, and can as a result adjust his gait, thus in Equation (4), $A$ quantifies the influence of ship swaying amplitude $B(t)$, while $\theta$ represents his gait cycle, which can be determined by,
\begin{eqnarray}
A = f(B) = F_s \times cos\psi = \beta B(t) cos\psi
\label{eq:05}
\end{eqnarray}
\begin{eqnarray}
\frac{d\theta}{dt} = \Omega + c B(t) sin(\varphi -\theta +\alpha)
\label{eq:06}
\end{eqnarray}
Here, $c$ quantifies pedestrians' sensitivity to ship swaying amplitude $B(t)$ and phase $\varphi(t)$. $\Omega$ is a random step frequency, which can be estimated following the pedestrian movement experiments \cite{Fang2012}. $\alpha$ is a phase lag parameter. $\beta$ represents the effect of friction when a pedestrian walks on a inclining ground.

\section{Results and discussion}
\label{sec:3}
The Equations (\ref{eq:01}) to (\ref{eq:06}) in Section \ref{sec:2} together describes forced movement pattern for a pedestrian on a swaying ship. It should be noticed that in reality, the ship swaying induced force component $F_s$ for a pedestrian might be very complex. So in the present section, we explore its effects by performing two simple case studies.
%
\begin{figure}[b]
\sidecaption
\includegraphics[scale=.5]{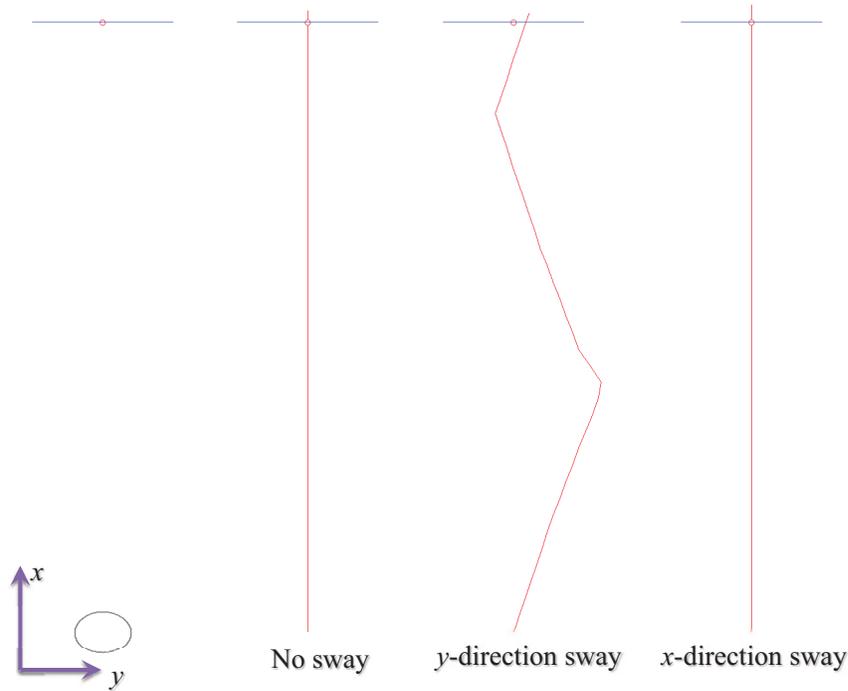}
%
%
\caption{Single pedestrian movement trajectories under different conditions.}
\label{fig:3}       
\end{figure}

\begin{figure}[b]
\sidecaption
\includegraphics[scale=.25]{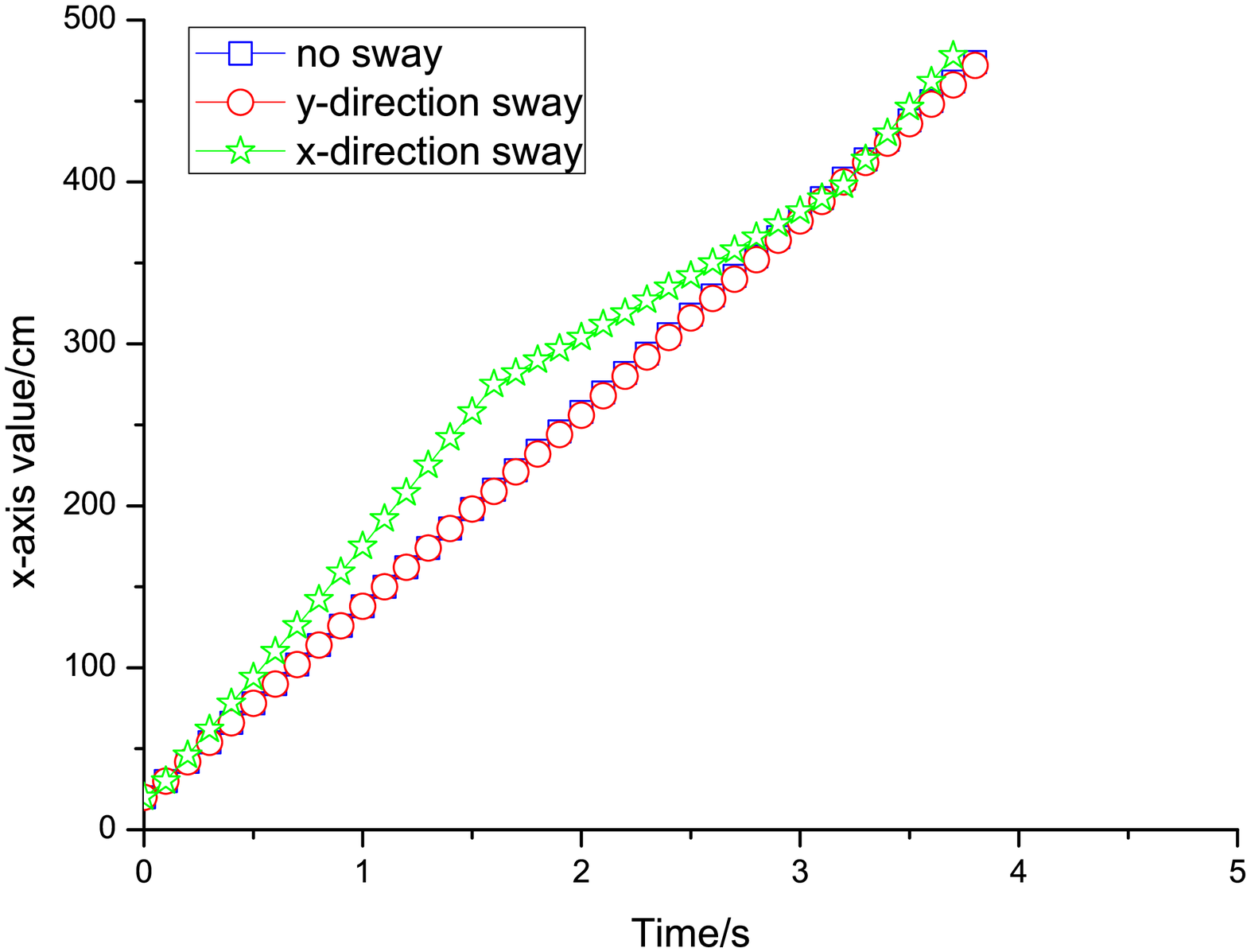}
\includegraphics[scale=.25]{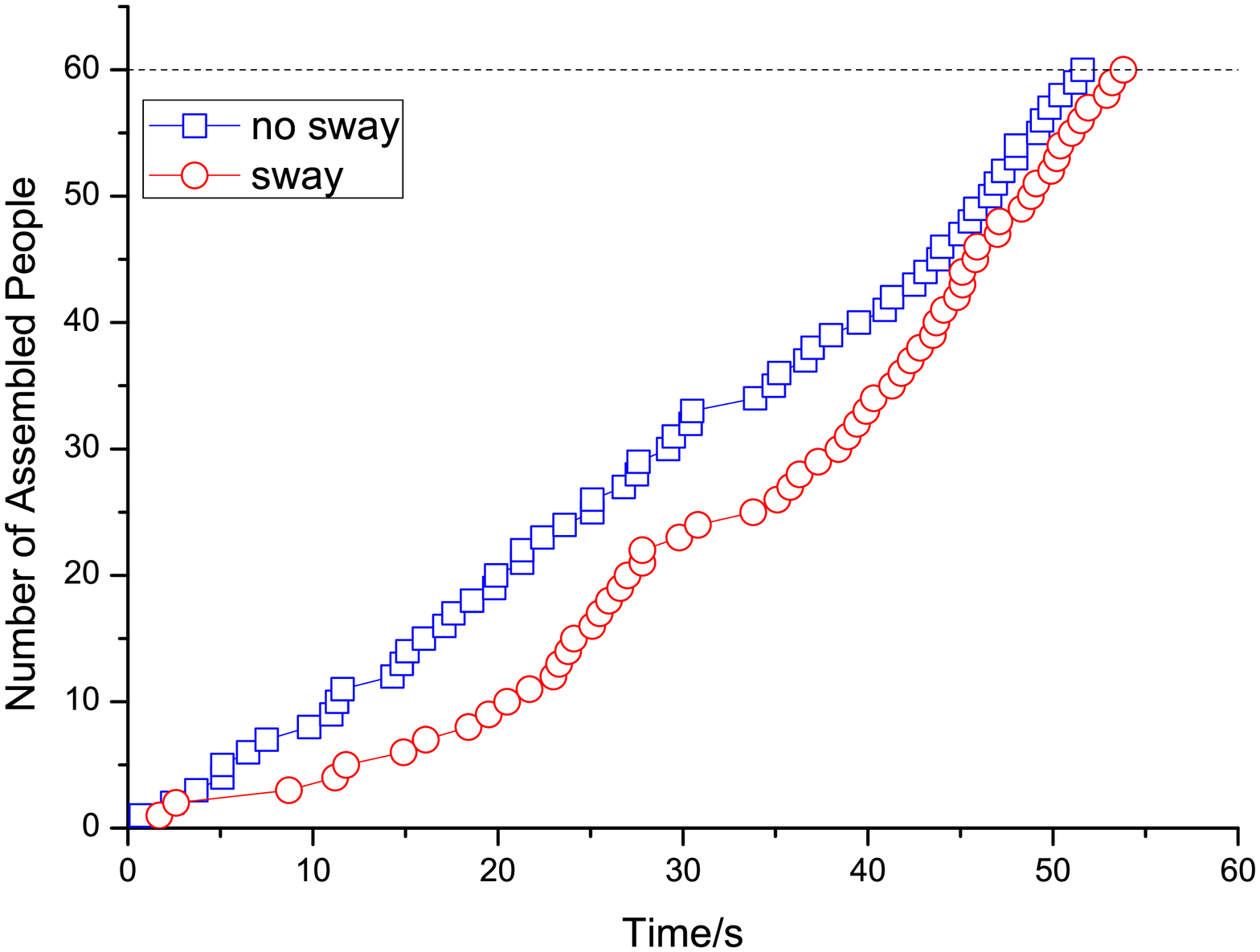}
%
%
\caption{Comparison of (a) pedestrian movement features and (b) ship passenger evacuation process.}
\label{fig:4}       
\end{figure}

Firstly, we place a pedestrian $5m$ away from his target, as shown in Fig.~\ref{fig:3}a. The pedestrian has a free movement speed of $1.2m*s^{-1}$. Three scenarios were considered. I) We assume the pedestrian is moving on a horizontal ground, and there is no ship swaying. II) The ship was swaying along the direction perpendicular to the pedestrian's movement direction. III) The ship is swaying along the direction which parallels to the pedestrian movement direction. Simulation snapshots for these three different scenarios can be found in Fig.~\ref{fig:3}b,~\ref{fig:3}c and~\ref{fig:3}d, respectively. As can be found in these figures, when there is no swaying, the pedestrian can move freely towards his target. When the ship was swaying perpendicular to the pedestrian movement direction, we can see as a result of the swaying, the pedestrian makes lateral movement when approaching his target. That is due to the ship swaying induced inertial force exerted on that pedestrian and changed his gait. We can also find although the pedestrian made lateral movement, the movement towards his direction has barely been influenced, as can be found in Fig.~\ref{fig:4}a. When the ship is swaying along the direction paralleling to the pedestrian movement direction, as shown in Fig.~\ref{fig:3}d, his trajectory seems like the one when there was no ship swaying, as in Fig.~\ref{fig:3}a. We further compared the spatiotemporal feature of these trajectories shown in Fig.~\ref{fig:3}a and~\ref{fig:3}d, and found as shown in Fig.~\ref{fig:4}a, the pedestrian accelerate at first and then decelerate, meaning he keeps changing his speed during the process when he moves towards his target in the case when the ship is swaying along the direction paralleling to his movement direction. We can see that the ship swaying affected pedestrian microscopic features even in these simple conditions.

%

\begin{figure}[b]
\sidecaption[t]
\includegraphics[scale=.5]{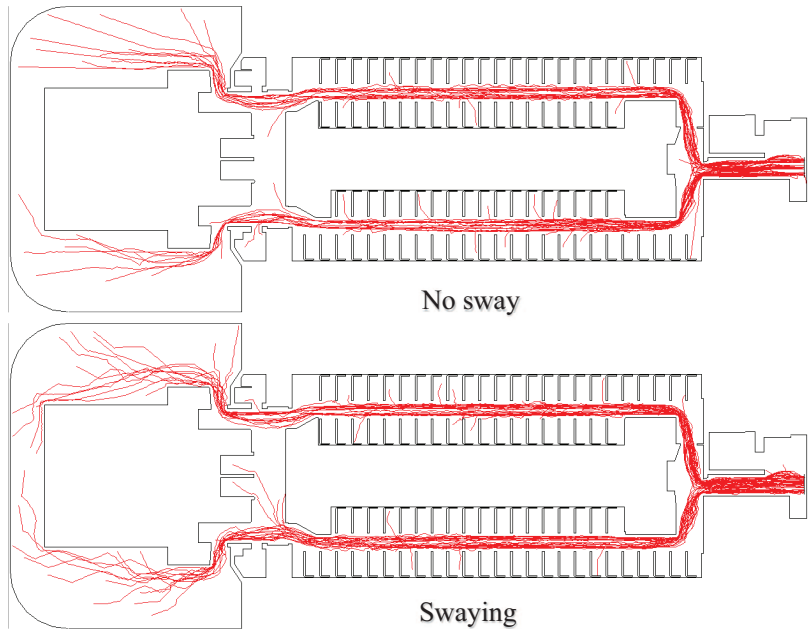}
%
%
\caption{Passenger evacuation simulation snapshots (a) with and (b) without ship swaying.}
\label{fig:5}       
\end{figure}

Secondly, scenarios of a ship evacuation with/without swaying were simulated. Considering that in case of emergency, prior to any decision to actually abandon the ship, passengers on a ship have to be evacuated to the assembling site. That is because if abandoning is unavoidable, these passengers can be evacuated immediately, and if there is no need to abandon ship, dealing with emergency would be much easier with no passengers around for those crew members. So, passengers were ordered to evacuate to the assembling area on the right side, i.e., front of ship in the present case, as shown in Fig.~\ref{fig:5}. There were in total 60 passengers on the ship. In Fig.~\ref{fig:5}a and~\ref{fig:5}b we show the trajectories of those passengers on the ship when there was no sway and there was sway, respectively. Comparing these figures we can easily find that due to ship swaying, passengers made lateral movements during evacuation. The lateral movement slowed down the evacuation process, as shown in Fig.~\ref{fig:4}a. We can also find that for those passengers initially located in the relatively open area, as shown in the left part of Fig.~\ref{fig:5}, their trajectories show clear curved features when the ship was swaying. When there was no sway, these trajectories were almost linear, representing these passengers can move freely towards their targets. For those who were located in the long channel, as can be found in the right part of Fig.~\ref{fig:5}, they were barely influenced by the swaying ship. The reason is that the channel is so narrow that passengers' lateral movements were hindered by those chairs and walls.

When we compare the evacuation process, as can be found in Fig.~\ref{fig:4}b, when there was no ship swaying, the total assembling time for 60 passengers is only slightly shorter than when there was ship swaying. That is because the evacuation processes were mainly performed in the long channel section, as shown in Fig.~\ref{fig:5}. But the assembling efficiency when there was no sway is always higher, as indicated by the number of assembled people shown in Fig.~\ref{fig:4}b. That is because, passengers located on the back and front sections had to make movement along the ship swaying direction, which would slower them down, as shown in Fig.~\ref{fig:4}a.
\section{Conclusion}
\label{sec:4}
An agent-based passenger evacuation model was built. Each agent in the model can sense the ship swaying feature and adjust its own gait, in this way, forced pedestrian movement feature as a result of ship swaying was mathematically quantified. Simulations of single pedestrian movement show that the angle between pedestrian movement direction and ship swaying direction may influence spatiotemporal features of a pedestrian. Thus under the situation of total evacuation, the passenger assembling efficiency would be affected. The simulated assembling time provides a very important criterion to evaluate the total evacuation time needed for a completed orderly evacuation. It should also be noticed that the computed evacuation time gives an estimate of the time the passengers need to get out of the ship interior and reach a boarding site, thus the influence of the ship interior can be evaluated to find out bottlenecks and make improvement accordingly.

\begin{acknowledgement}
The authors sincerely appreciate the supports from the Research Grant Council of the Hong Kong Administrative Region, China (Project No. CityU11209614), and from China National Natural Science Foundation (No. 71473207, 51178445, 71103148) as well as the Fundamental Research Funds for the Central Universities (2682014CX103).
\end{acknowledgement}
%

\bibliographystyle{spmpsci}
\bibliography{tgfref1}

\end{document}